\newcommand{\HI}{H\,{\sc {i}}}
\newcommand{\HII}{H\,{\sc {ii}}}

\newcommand{\rmaj}{$R_{\rm maj}$}

\newcommand{\lsim}{~\rlap{$<$}{\lower 1.0ex\hbox{$\sim$}}}
\newcommand{\gsim}{~\rlap{$>$}{\lower 1.0ex\hbox{$\sim$}}}

\documentclass[twocolumn]{aastex62}

\usepackage{multirow}
\usepackage{gensymb}

\received{February 4, 2020}
\accepted{February 19, 2020}
\toappearin{ApJL}

\shorttitle{Dwarf-dwarf galaxy merger VCC848}
\shortauthors{Zhang et al.}

\begin{document}

\title{The Blue Compact Dwarf Galaxy VCC848 Formed by Dwarf-Dwarf Merging}

\author[0000-0003-1632-2541]{Hong-Xin Zhang}
\affil{CAS Key Laboratory for Research in Galaxies and Cosmology, Department of Astronomy, University of Science and Technology of China, Hefei, Anhui 230026, China}
\affil{School of Astronomy and Space Science, University of Science and Technology of China, Hefei 230026, China}
\email{hzhang18@ustc.edu.cn}

\author[0000-0003-2922-6866]{Sanjaya Paudel}
\affil{Department of Astronomy and Center for Galaxy Evolution Research, Yonsei University, Seoul 03722}
\email{sanjpaudel@gmail.com}

\author[0000-0001-5303-6830]{Rory Smith}
\affil{Korea Astronomy and Space Science Institute, Daejeon 305-348, Republic of Korea}
\email{rorysmith@kasi.re.kr}

\author{Pierre-Alain Duc}
\affil{Université de Strasbourg, CNRS, Observatoire astronomique de Strasbourg, UMR 7550, F-67000 Strasbourg, France}

\author{Thomas H. Puzia}
\affil{Instituto de Astrofísica, Pontificia Universidad Católica de Chile, 7820436 Macul, Santiago, Chile}

\author{Eric W. Peng}
\affil{Department of Astronomy, Peking University, Beijing 100871, China}
\affil{Kavli Institute for Astronomy and Astrophysics, Peking University, Beijing 100871, China}

\author{Patrick C\^ote}
\affil{National Research Council of Canada, Herzberg Astronomy and Astrophysics Program, 5071 West Saanich Road, Victoria, BC V9E 2E7, Canada}

\author{Laura Ferrarese}
\affil{National Research Council of Canada, Herzberg Astronomy and Astrophysics Program, 5071 West Saanich Road, Victoria, BC V9E 2E7, Canada}

\author{Alessandro Boselli}
\affil{Aix Marseille Université, CNRS, LAM (Laboratoire d’Astrophysique de Marseille) UMR 7326, F-13388 Marseille, France}

\author{Kaixiang Wang}
\affil{Department of Astronomy, Peking University, Beijing 100871, China}
\affil{Kavli Institute for Astronomy and Astrophysics, Peking University, Beijing 100871, China}

\author{Se-Heon Oh}
\affil{Department of Physics and Astronomy, Sejong University, 209 Neungdong-ro, Gwangjin-gu, Seoul, Republic of Korea}

\begin{abstract}
It has long been speculated that many starburst or compact dwarf galaxies are resulted from dwarf-dwarf galaxy merging, but unequivocal  
evidence for this possibility has rarely been reported in the literature.\ We present the first study of deep optical broadband images of a gas-dominated 
blue compact dwarf galaxy (BCD) VCC848 (M$_{\star}$$\simeq$2$\times$10$^{8}$M$_{\odot}$) which hosts extended stellar shells and thus 
is confirmed to be a dwarf-dwarf merger.\ VCC848 is located in the outskirts of the Virgo Cluster.\ By analyzing the stellar light distribution, 
we found that VCC848 is the result of a merging between two dwarf galaxies with a primary-to-secondary mass ratio $\lesssim$ 5 for the stellar 
components and $\lesssim$ 2 for the presumed dark matter halos.\ The secondary progenitor galaxy has been almost entirely disrupted.\ The age-mass 
distribution of photometrically selected star cluster candidates in VCC848 implies that the cluster formation rate (CFR, $\propto$ star formation rate) was 
enhanced by a factor of $\sim$ $7-10$ during the past $\sim$ 1 Gyr.\ The merging-induced enhancement of CFR peaked near the galactic center a few hundred 
Myr ago and has started declining in the last few tens of Myr.\ The current star formation activities, as traced by the youngest clusters, mainly occur at large 
galactocentric distances ($\gtrsim$ 1 kpc).\ The fact that VCC848 is still (atomic) gas-dominated after the period of most violent collision suggests that gas-rich 
dwarf galaxy merging can result in BCD-like remnants with extended atomic gas distribution surrounding a blue compact center, in general agreement with 
previous numerical simulations.

\end{abstract}


\section{Introduction} \label{sec:intro}

Galaxy mergers, especially those involving gas-rich comparable-mass galaxies, can dramatically change galaxy morphologies, enhance star 
formation activities and trigger active galactic nuclei in short timescales \citep[][and references therein]{naab17}.\ The majority of studies of 
galaxy mergers so far have focused on relatively massive galaxies, while mergers between dwarf galaxies (M$_{\star}$ $<$ 10$^{9}$) received 
little attention until very recently.\ In contrast to their massive counterparts, star-forming dwarf galaxies, and hence their mergers, are often gas 
dominated.\ Relative proportion of the dissipative gas component and non-dissipative stellar component can make substantial differences to 
the merging process.\

The first solid evidence for a dwarf-dwarf merger was found in the dwarf spheroidal galaxy Andromeda II by \cite{amorisco14}.\
Since then \citep[see however][]{chilingarian09}, signatures of past merger events have been reported in several early-type dwarf 
galaxies \citep[][]{paudel17a,cicuendez18}.\ \cite{stierwalt15} carried out the first systematic study of gas-rich dwarf-dwarf interacting 
pairs and found that star formation rate (SFR) of relatively close pairs is enhanced by a factor of $\sim$ 2 on average.\ \cite{pearson16} 
further showed that gas-rich dwarf pairs have more extended atomic gas distribution than unpaired analogues.\ In addition, case studies 
of several more star-forming dwarf pairs have been carried out recently \citep{annibali16, privon17, paudel17b, paudel18b, makarova18, 
johnston19}.\

Previous studies of gas-rich dwarf-dwarf interactions are biased towards well-separated pairs, which is partly due to the difficulty in identifying  
merging signatures from morphologically irregular dwarf galaxies in general.\ It remains to be seen how dwarf-dwarf merger impacts the overall 
star formation activities and morphology of merger remnants.\ Here we present a study of a unique blue compact dwarf galaxy (BCD), VCC848 
($M_{B}=-16.05$ mag; \citealt{gildepaz03}), which stands out with faint but remarkably extended stellar shells wrapping around the stellar main 
body.\footnote{After a systematic search for fine substructures in Virgo star-forming dwarfs, we found several more Virgo BCDs ($\sim$ 25\%) with 
hints of tidal shells, but they are not as impressive as those in VCC848.} Stellar shells are unambiguous signatures of minor or major galaxy merger 
events in the recent past \citep[e.g.][]{hernquist88, hernquist92a}, but have not been reported for star-forming dwarfs previously.\ 

\begin{figure*}
\centering
\includegraphics[width=\textwidth]{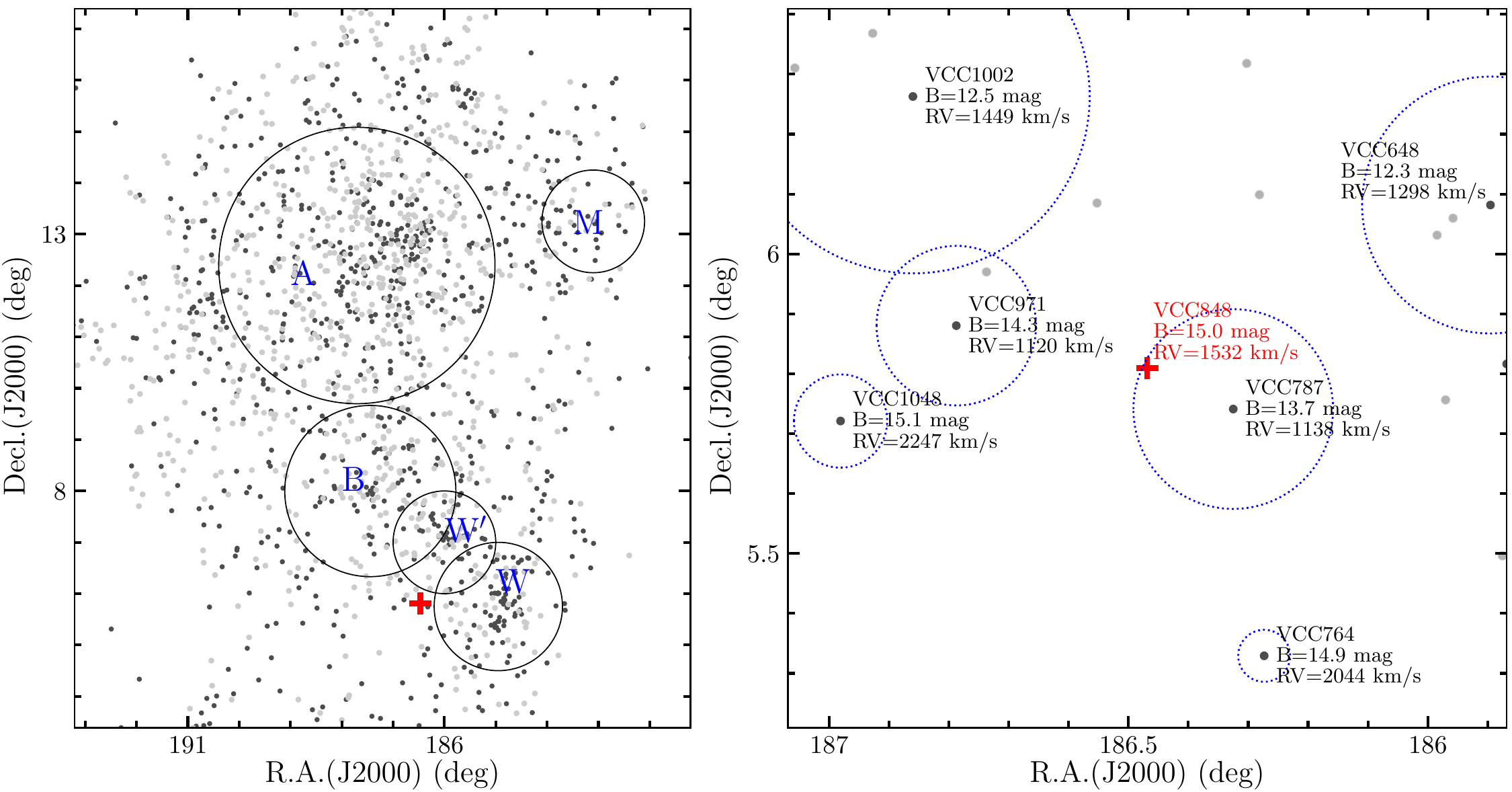}
\caption{
Location of VCC848 in the sky area of the Virgo cluster.\ 
The left panel presents the spatial distribution of redshift-confirmed ({\it black symbols}) or candidate ({\it gray dots}) VCC galaxies according to 
the most recent compilation of redshift measurements by the NGVS team.\
The {\it red plus} symbol marks the location of VCC848.\ The two biggest black solid circles mark half the virial radius of M87 (A) and of 
M49 (B) sub-clusters respectively, and the three small black solid circles mark the boundaries of the M, W$^{\prime}$ and W clouds, as defined 
in \cite{binggeli87}.\ The right panel is a zoom-in of the 1.2$\times$1.2 degree sky area around VCC848.\ The blue dotted circles in the right panel 
mark half the virial radii of individual confirmed VCC galaxies, where the virial radius is approximated as the $i$ band half-light radius times 67 \citep{kravtsov13}. 
\label{fig_virgoradec}}
\end{figure*}

VCC848 (RA=186.46853$^{\degree}$, DEC=5.80930$^{\degree}$) is located in the outskirts of the Virgo cluster.\ As illustrated in Figure \ref{fig_virgoradec}, 
it is 6.7$^{\degree}$ ($\simeq$ 1.9 Mpc in projection) away from the Virgo central galaxy M87 and 2.4$^{\degree}$ ($\simeq$ 0.7 Mpc in projection) 
away from the dominant galaxy M49 of the Virgo B sub-cluster, and it is also outside of the boundaries of other Virgo sub-structures.\ It has a radial velocity of 
1532 km/s, which can be compared to $\sim$ 1000 km/s of the Virgo B sub-cluster.\ Furthermore, VCC848 is virtually free of gravitational influence from its 
neighboring galaxies, given its current location and radial velocity.\ VCC848 is rich in neutral \HI~gas ($M$(\HI)$=4.2\times10^{8}$$M_{\odot}$, \citealt{haynes11}), 
with $M$(\HI)/$L_{B}$ $\simeq$ 1.0 $M_{\odot}/L_{\odot}$, and has an \HI~deficiency parameter of $-0.2$ (\citealt{grossi15}), which means that its \HI~gas mass 
is 0.2 dex higher than that expected for isolated galaxies of similar Hubble types and sizes (\citealt{haynes84}).\ By using the SFR estimator from \cite{catalan15}, 
the current SFR of VCC848 is estimated to be 0.023 $M_{\odot}$/yr based on H$\alpha$ image from \cite{gildepaz03} and Herschel far-infrared photometry from 
\cite{grossi15}.\ According to the SFR$-$stellar mass relation \citep{shin19}, VCC848 is 0.19 dex above the star formation main sequence for its stellar mass 
(2$\times$10$^{8}$ $M_{\odot}$; see below).\ 

This paper aims to exploit the broadband imaging data of VCC848 to investigate the impact of merging on the stellar distribution and star formation history (SFH).\ 
In Section \ref{sec:obs}, we describe the data used in this work.\ In Section \ref{sec:ellipse}, we present the stellar light distribution and isophotal analysis.\ 
In Section \ref{sec:cluster}, we present the detection of star cluster candidates and its  implications for the past SFH.\ A summary is given in Section \ref{sec:sum}.\ 
Throughout this paper, the \cite{schlafly11} Galactic extinction map is used to correct our photometry and a distance of 16.5 Mpc is adopted for VCC848 \citep{blakeslee09}.

\section{Broadband optical imaging data} \label{sec:obs}

Broadband $u, g, i, z$ band images of VCC848 were obtained with the MegaCam instrument on the Canada-France-Hawaii Telescope.\    
The observations are part of the Next Generation Virgo Cluster Survey (NGVS; \citealt{ferrarese12}), which reaches a 2$\sigma$ surface brightness limit of 
$\mu_{g}\simeq29$ mag arcsec$^{-2}$.\ The processed NGVS images for the VCC848 field have 0\farcs186 pixel scales and Point Spread Function Full Width 
at Half Maximum (FWHM) of 0\farcs78, 0\farcs71, 0\farcs53 and 0\farcs62 respectively for $u$, $g$, $i$ and $z$ passbands.\ 

\section{Stellar light distribution}
\label{sec:ellipse}

\subsection{Description of the tidal features}
\begin{figure*}[tbp]
\centering
\includegraphics[height=0.43\textheight]{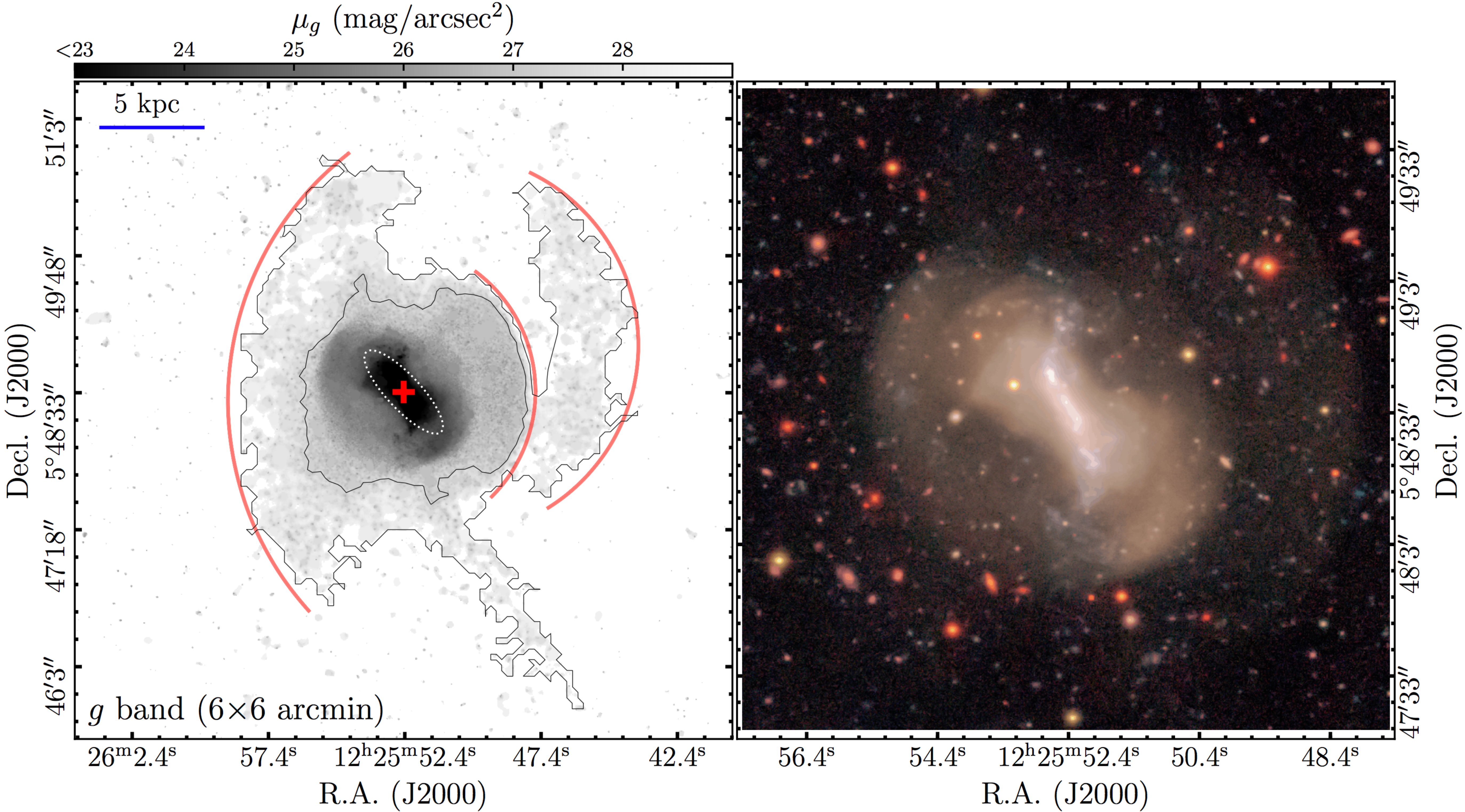}
\caption{
{\it Left: }$g$ band image of VCC848.\ The original NGVS image has been adaptively smoothed to a minimum S/N of 
5.\ The two contour levels mark, respectively, a surface brightness of 27.3 and 28.3 mag arcsec$^{-2}$.\ Three most obvious 
sharp-edged shell-like structures are marked by orange curves.\ The photometric center of VCC848 is marked by a red plus 
symbol.\ The white dotted ellipse has $R_{\rm maj}$=30\arcsec~and encloses the stellar main body dominated by the primary 
progenitor of the merger.\ {\it Right: }Central 2.5$\times$2.5 arcmin of the $u,g,i$ color composite image.\ 
\label{fig_g}}
\end{figure*}

In order to better visualise faint substructures, we mask out bright foreground stars or background galaxies in the $g$ image, 
and then use the adaptive smoothing code {\sc adaptsmooth} (\citealt{zibetti09}) to adaptively smooth the masked $g$ band image to a 
minimum S/N of 5/pixel.\ The smoothed image is shown in Figure \ref{fig_g}.\ One can notice three extended shell-like structures around 
the bright stellar main body.\ These shell structures are largely aligned along the east-west direction, and the innermost one is brighter and 
has sharper edges than the two at larger radii.\ Moving toward smaller radii, there is a circularly-edged stream-like feature wrapping nearly 
180\degree~around the stellar main body to the southeast.\ Lastly, there is a straight stream-like narrow feature which extends to the southwest 
and is largely parallel to the orientation of the stellar main body.

\subsection{Isophotal analysis}
\begin{figure*}[tbp]
\centering
\includegraphics[width=0.99\textwidth]{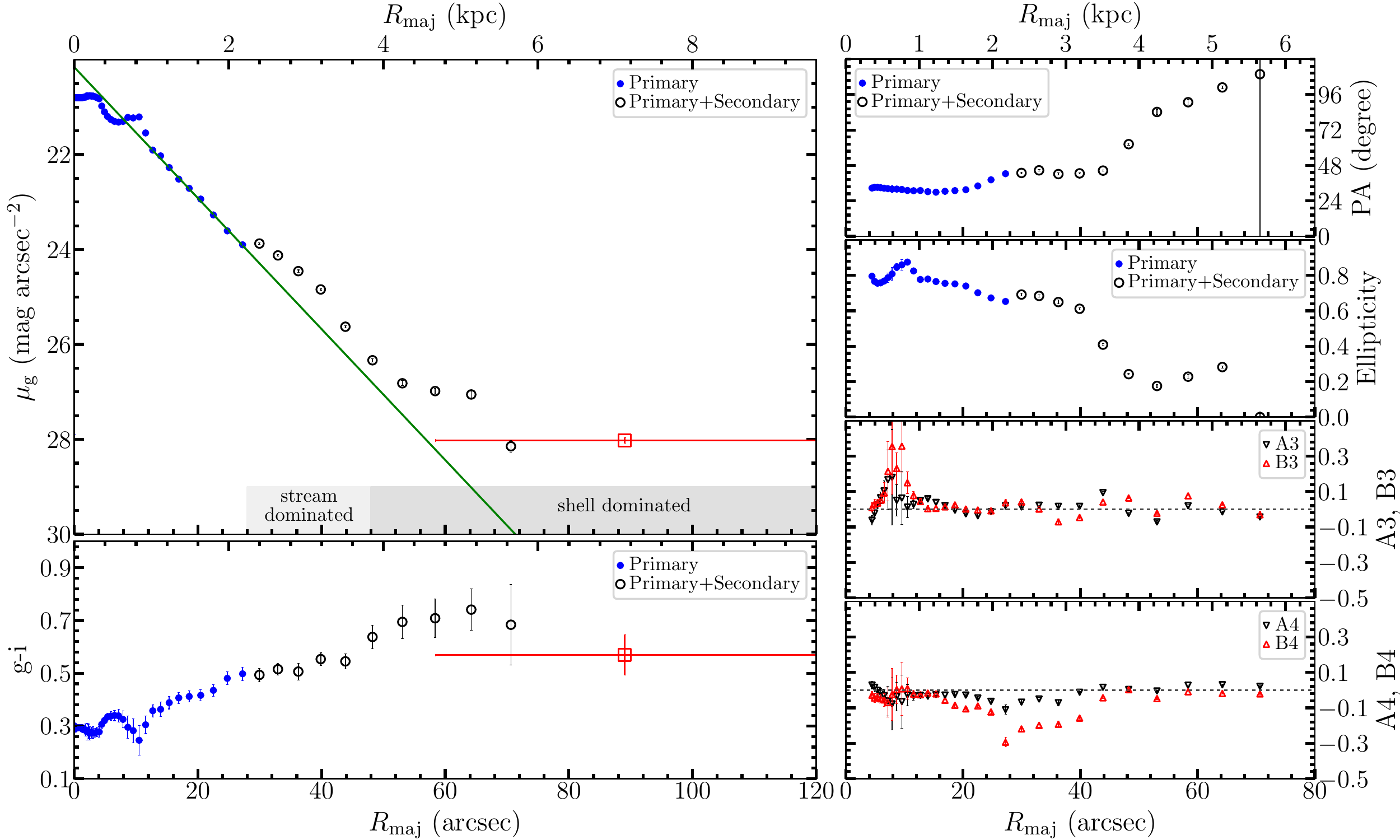}
\caption{
Results of {\sc ellipse} surface photometry of VCC848.\
The upper left panel shows the $g$ band surface brightness profile, and the lower left panel shows the ($g-i$) color profile.\
Radial variations of other {\sc ellipse} isophote parameters are shown in the right panels.\ When plotting the surface brightness, 
color, PA and ellipticity profiles, the inner radial ranges that are dominated by the primary galaxy are plotted 
as {\it blue filled circles} while the outer radial ranges with significant contribution from both the primary and secondary as {\it black  
open circles}.\ The {\it red open squares} in the {\it left} panels represent the photometry for the irregular region enclosed 
by the 27.3 and 28.3 mag arcsec$^{-2}$ $g$ band contour levels as marked in Figure \ref{fig_g}.
\label{fig_ell}}
\end{figure*}

To quantify the stellar light distribution, we perform surface photometry in the original $g$ and $i$ band images with the Image Reduction 
and Analysis Facility (IRAF) task {\sc ellipse}.\ Our isophotal fitting is carried out in an iterative manner in order to find the photometric center, 
which is constant with radius, and the ellipticity and position angle (PA) which vary with radius.\ The results are shown in Figure \ref{fig_ell}.\

The stellar main body of VCC848 has twisted isophotes, with the PA varying from $\simeq$ 32$\degree$ near the center to 42$\degree$ at 
a major axis radius \rmaj~$\simeq$ 28$\arcsec$ ($\sim$ 2.2 kpc) where boxy distortion of the isophotes reaches the maximum 
({\sc ellipse} B4 coefficient $\simeq$ $-0.3$; bottom right panel of Figure \ref{fig_ell}).\ Beyond the central $\sim$ 30\arcsec~(i.e.\ the 
stellar main body) are the above-described tidal streams and shells, which apparently drive the abrupt decrease of isophotal ellipticities at 
large radii (\rmaj~$\gtrsim$ 40\arcsec).\ Photometry integrated over the area enclosed by the 27.3 and 28.3 mag arcsec$^{-2}$ 
$g$ band brightness contours is shown as red open square symbols in Figure \ref{fig_ell}.\

The ($g-i$) colors become redder at larger radii.\ The obscured star formation traced by Herschel far-infrared photometry accounts for 
merely 9\% of the SFR budget of VCC848 (Section \ref{sec:intro}).\ So the radial color trend primarily reflects an increasingly older 
stellar population toward larger radii.\ The shell-dominated region reaches $g-i \simeq 0.6-0.7$ mag which, for single stellar populations 
(SSP) at one-fifth solar metallicity (i.e.\ the nebular gas abundance measured by \citealt{vilchez03}), would correspond to stellar ages of $2-3$ Gyr.\ 

\subsection{Implications for the primary-to-secondary mass ratio and collision geometry}

The radial surface brightness distribution at \rmaj~$\lesssim$ 28$\arcsec$ follows an exponential profile, suggesting a disc structure which 
is presumably dominated by the primary progenitor of this merging system.\ The best-fit exponential profile for the disc-dominated 
radial range is over-plotted as a blue straight line in Figure \ref{fig_ell}.\ To make an estimate of the respective stellar luminosities of the 
primary and secondary galaxies of the merging system, we assume that the region within \rmaj~$=$ 28$\arcsec$ is contributed exclusively 
by the primary, while the region beyond has significant contribution from both the primary and secondary.\ The total luminosity of the primary 
is a sum of the luminosity directly measured in the images within \rmaj~$=$ 28$\arcsec$ and an exponential extrapolation beyond by using 
the average {\sc ellipse} geometric parameters at 26$\arcsec$ $<$ \rmaj~$<$ 28$\arcsec$.\ The luminosity of the secondary is calculated by 
subtracting the luminosity of the primary from the total luminosity enclosed within the 28.3 mag arcsec$^{-2}$ contour.\

The above exercise of decomposing the merging system gives a $g$ band absolute magnitude $M_{g}$ of $-16.28$ and a ($g-i$) color of 0.45 
for the primary, and a $M_{g}$ of $-14.39$ and a ($g-i$) color of 0.53 for the secondary.\ To estimate the stellar mass, we first transform our 
MegaCam magnitude to the SDSS magnitude using the relation: ($g-i$)$_{\rm SDSS}$ $=$ 1.1$\times$($g-i$)$_{\rm MegaCam}$ and $i_{\rm SDSS}$ 
$=$ $i_{\rm MegaCam}$ + 0.005$\times$($g-i$)$_{\rm MegaCam}$\footnote{The relations are derived based on the best-fit stellar population models 
for Local Group dwarf galaxies presented in \cite{zhang17}}, and then obtain mass-to-light ratios using the color--mass-to-light relation calibrated for  
Local Group dwarf galaxies by \cite{zhang17} based on the FSPS stellar population models \citep{conroy09}.\ The resultant stellar masses are 1.7$\times$10$^{8}$ 
$M_{\odot}$ and 3.8$\times$10$^{7}$ $M_{\odot}$ for the primary and secondary respectively, leading to a $4.5:1$ primary-to-secondary stellar mass ratio.\ 
By adopting the stellar-to-dark matter halo mass relation from \cite{guo10}, the corresponding dark matter halo masses for the primary and secondary 
are 5.6$\times$10$^{10}$ and 3.4$\times$10$^{10}$ $M_{\odot}$ respectively, giving a halo mass ratio of $1.6:1$.\ The mass ratio estimated here is  
likely an upper limit if allowing for non-negligible contribution of the secondary to the central exponential part of the system.\

The fact that we observe mostly extended stellar shells, rather than tidal tails, around the merging system favors a nearly radial encounter \citep{quinn84}.\ 
The overall alignment of the stellar shells suggests that the collision was largely along the east-west direction.\ Because we do not see any distinct 
stellar concentrations other than the main body, the secondary progenitor has been almost entirely disrupted, leaving behind the observed stellar 
shells.\ 

\section{Star clusters and their implications for star formation history}\label{sec:cluster}

\subsection{Detection and aperture photometry}
We use {\sc SExtractor} \citep{bertin96} to detect star cluster candidates in the original $i$ band image which has the highest spatial resolution among our images.\ 
Ordinary star clusters ($r_{\rm eff}$ $\sim$ 2$-$10 pc; e.g.\ \citealt{portegies10}) at the distance of VCC848 are expected to be point-like or marginally 
resolved sources at our $i$ band resolution \citep[e.g.][]{durrell14}.\ We run {\sc SExtractor} with a BACK\_SIZE parameter of 9 pixels and a detection 
threshold of 3 times the rms noise.\ A total of 1994 sources are detected in a 7$\farcm$2$\times$11$\farcm$2 cutout image centered around VCC848.\
By following the routine procedure of completeness estimation based on artificial star tests \citep[e.g.][]{munoz14}, we find that our detection 
reaches a 90\% completeness limit at $i$ $=$ 23.5 mag.\ 1078 of the 1994 sources have $i$ $\leq$ 23.5 mag.\ 

We perform $i$-band aperture photometry for the detected sources with a 5-pixel (0.93$\arcsec$) diameter aperture size.\
The aperture size is chosen to contain more than half of the total flux expected for a point source and at the same time to reduce as much as 
possible the contamination from neighboring sources.\ Local background for each source is determined using a 4-pixel wide background annulus, 
with an inner radius of 6 pixel.\ We apply a multiplicative correction factor of 1.8 to the background-subtracted aperture photometry to recover the 
expected total flux for $i$-band point sources.\ To determine the colors, we smooth the $g,i,z$ images to match the $u$-band spatial 
resolution, and follow the same procedure described above to obtain $u,g,i,z$ photometry based on the resolution-matched images.\ 


\subsection{Selection of star cluster candidates in VCC848}
We select star cluster candidates taking advantage of both the shape and color information.\ In particular, we use the criteria of ELLIPTICITY 
$<$ 0.2 and FWHM $<$ 4.5 pixels to pick out point-like or marginally-resolved objects with $i < 23.5$ mag.\ The {\sc SExtractor} shape parameter 
measurements are subject to large uncertainties in regions with steep background gradient, so we relax the requirement for shape parameters if a 
source in question is spatially coincident an \HII~region visible in the H$\alpha$ image of \cite{gildepaz03}.\ The above selection criteria leave us 
with 526 sources to be considered for further selection with colors.\ 

\begin{figure*}[p]
\centering
\includegraphics[width=0.99\textwidth]{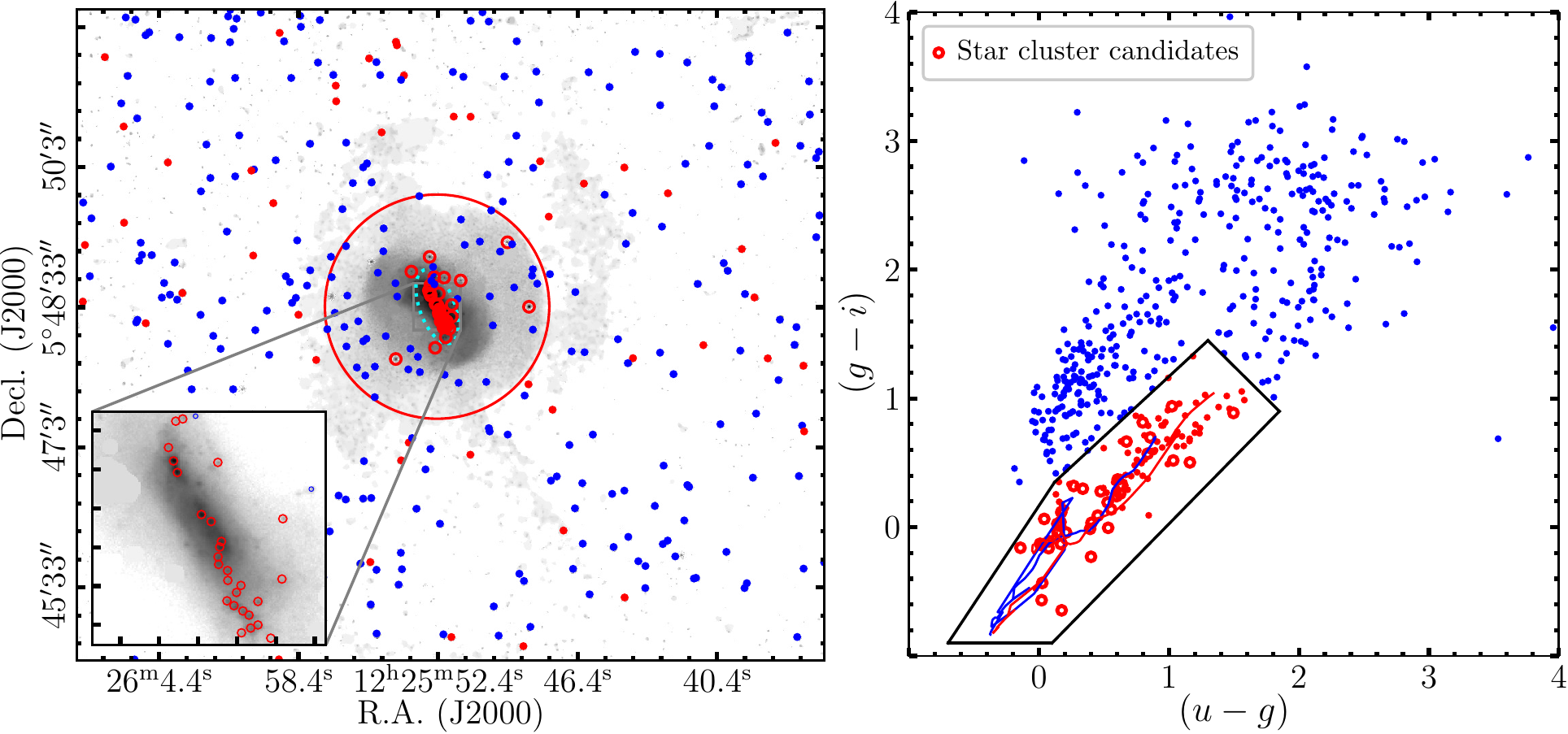}
\caption{
Star cluster candidates with $i$ $<$ 23.5 mag detected in the VCC848 field.\ 
The {\it left} panel shows the central 7$\arcmin$$\times$8$\arcmin$ of the adaptively smoothed $g$ band image and the inset figure to the bottom left 
is the central 30$\arcsec$$\times$30$\arcsec$ of the original $i$ band image.\ The blue dots represent the point or marginally-resolved sources selected 
based on $i$-band shape parameters (see the text for details), the red dots represent the point or marginally-resolved sources that fall within a ($u-g$) 
vs.\ ($g-i$) parameter space occupied by star clusters, as delineated by a closed black polygon in the {\it right} panel, and the small red open circles 
represent our final sample of 37 star cluster candidates that satisfy the shape and color criteria and are also within a 1.2\arcmin-radius circular region 
around VCC848 (big red circle in the {\it left} panel).\ For comparison purpose, the FSPS single stellar population models of $\log(Z/Z_{\odot})$ = 
$-1.98$ and $-0.39$ \citep[ages from 0.3 Myr to 15 Gyr;][]{conroy09} are overplotted as blue and red curves respectively in the right panel.\
The cyan dotted ellipse in the left panel marks the area used for completeness estimates.\ 
\label{fig_clusterugi}}
\end{figure*}

The right panel of Figure \ref{fig_clusterugi} shows the ($u-g$) vs.\ ($g-i$) distribution of the above-selected 526 sources.\ The sources 
occupy two distinct branches.\ The upper branch is identified to be background galaxies, while the lower branch is largely occupied by 
star clusters and foreground stars.\ We draw a polygon to delineate a region in the color-color space to single out the most probable 
star cluster candidates (116 in total).\ Lastly, we require that the VCC848 star cluster candidates be located within a 1.2\arcmin-radius 
circular region, approximately matching the area enclosed by the 27.3 mag arcsec$^{-2}$ $g$ band brightness contour.\ This last step leaves 
us with a final sample of 37 star cluster candidates.\

The average number density of the 79 ($116-37$) color-selected candidates located outside of the 1.2\arcmin-radius circular region around VCC848 
is $\simeq$ 1.0 arcmin$^{-2}$.\ Therefore, we expect an average of $\simeq$ 4.5 contaminants for the 1.2\arcmin-radius circular region, giving a $\simeq$ 88\% 
purity for our star cluster sample.\ We note, however, that 33 of the 37 candidates are concentrated within the central 30\arcsec, implying a 98\% purity.

\subsection{Stellar population modeling}
We estimate stellar ages and masses of our star cluster candidates by fitting the FSPS SSP models to $u,g,i,z$ photometry.\ Considering the age-metallicity 
degeneracies, we perform two sets of extinction-free stellar population fitting.\ One imposes a Gaussian prior constraint on metallicities, with a mean of $-0.88$ 
dex (chosen to be 0.2 dex lower than the abundance measurements for \HII~regions by \citealt{vilchez03}) and a standard deviation of 0.3 dex, while the other 
one does not impose prior constraints on metallicities.\ An upper limit of one-third solar metallicity is adopted for both sets of fitting.\ We will show that the two 
sets of spectral energy distribution (SED) fitting give statistically similar age-metallicity distributions.\ With the best-fit age and metallicity of each cluster, we infer 
the stellar mass at birth by correcting the present-day mass for mass loss due to stellar evolution, based on the FSPS models.\ As we will show in Figure \ref{fig_agemass}(c), 
mass loss caused by dynamical evaporation is not expected to be important for our cluster sample.

\subsection{Age-mass relation and its implication for star formation history of VCC848}
The stellar masses at birth are plotted against ages of our star cluster candidates in Figure \ref{fig_agemass}(c).\ The lower limit of cluster mass 
increases with age, which is jointly driven by the luminosity-limited detection efficiency, evolutionary fading and dynamical evaporation.\ 
The maximum cluster mass also varies with age, which may be driven by the size-of-sample effect, a time-dependent cluster formation 
rate (CFR) or cluster mass function (CMF).\ CMF of young star clusters observed in nearby galaxies can always be described by a power law 
$dN/dM$ $\propto$ $M^{\alpha}$, where $\alpha$ $\simeq$ $-2\pm0.2$ \citep[see][and references therein]{krumholz19}.\ For a constant 
CFR and $\alpha = -2$ power law CMF, the maximum cluster mass per logarithmic time interval scales linearly with age due to the size-of-sample 
effect \citep[e.g.][]{hunter03}.\ Such a linear $M-$age relation is illustrated in Figure \ref{fig_agemass}(c).

Given that dynamical evaporation and tidal shocking are not expected to significantly influence star clusters with M$_{\star} > 10^{5}$ $M_{\odot}$ 
\citep{reinacampos18}, the CFR of VCC848 has been significantly enhanced during the past $\sim$ 1.0 Gyr, because the observed maximum cluster 
masses above 10$^{5}$ $M_{\odot}$ do not rise steadily with lookback time, which would otherwise be expected for the size-of-sample effect.\ However, 
during the last $\sim$ $20-30$ Myr,  
the CFR appears to have started declining.\ The brightest star cluster of VCC848, with a $V$ band absolute magnitude $M_{V}^{\rm brightest}$ $=$ $-11.15$ 
mag (derived based on the best-fit model SED), is $>$ 3$\sigma$ above the average $M_{V}^{\rm brightest}$$-$SFR relation for nearby galaxies 
(Figure \ref{fig_agemass}(b)), in line with a recent decline of SFR \citep{bastian08}.\

Our cluster sample is $\gtrsim$ 90\% complete at M$_{\star} \geq$ 10$^{5.25}$ $M_{\odot}$ across nearly the whole age range.\ 
Figure \ref{fig_agemass}(d) shows the average number of clusters with M$_{\star}$ $>$ 10$^{5.25}$ $M_{\odot}$ over age intervals of 0$-$1 Gyr and 
1$-$13 Gyr.\ The average CFR in the last 1 Gyr is $\sim$ $7-10$ times that at earlier times.\ If the mass fraction of stars born in star clusters is independent 
of SFR, as suggested by recent studies \citep[e.g.][]{chandar17}, the trend for CFR would be approximately equivalent to that for SFR.\

\subsection{Spatial distribution}
\begin{figure*}[p]
\centering
\includegraphics[width=0.99\textwidth]{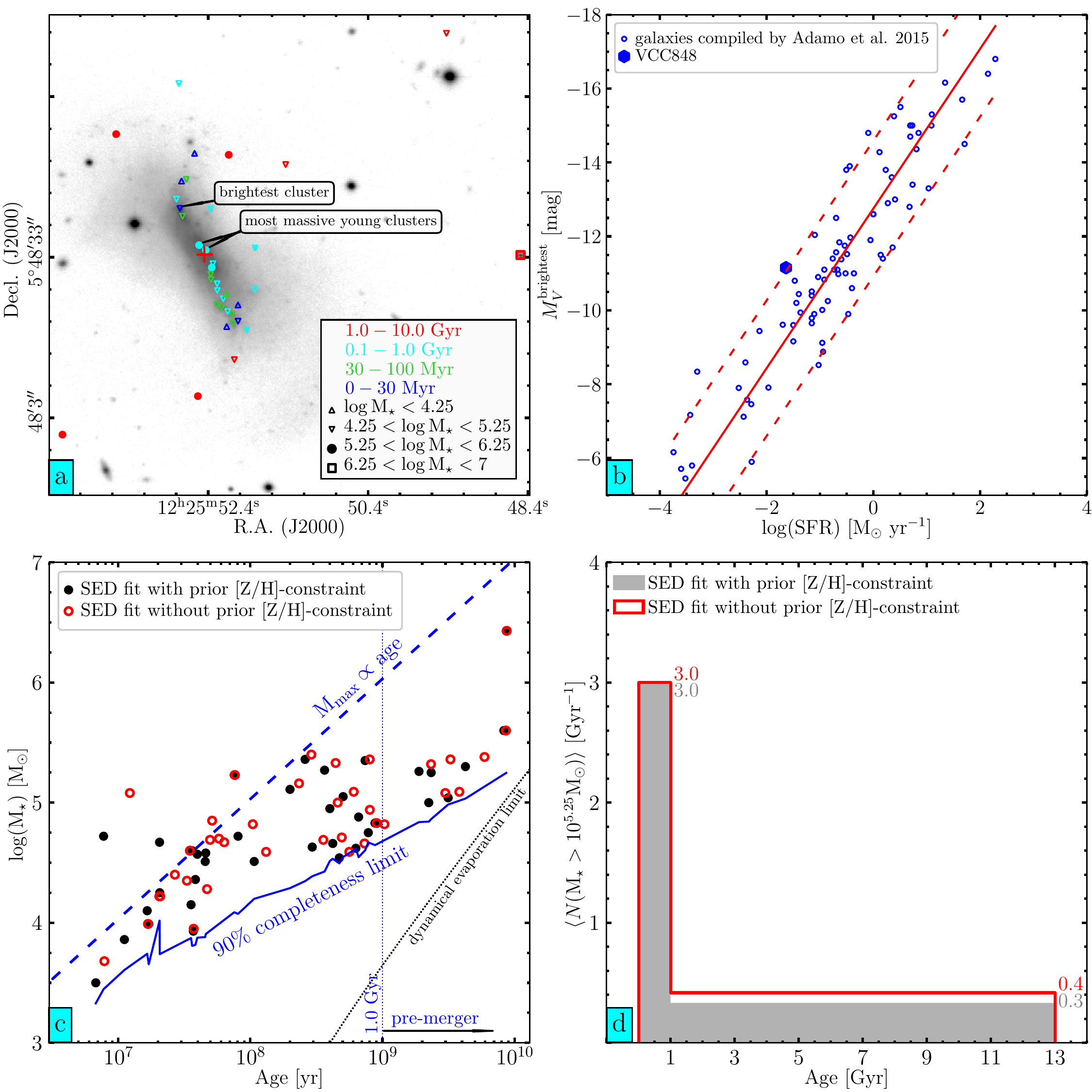}
\caption{
{\it Panel a}: Spatial distribution of the 37 star cluster candidates in VCC848.\
{\it Panel b}: The absolute magnitude of the brightest young star cluster is plotted against the SFR of its host galaxy.\ The blue open circles 
are nearby galaxies compiled by \cite{adamo15} and the blue filled hexagon represents VCC848.\ The red solid line is the best-fit linear relation 
($M_{V}^{\rm brightest}$ $= -2.16\log({\rm SFR}) + 12.76$) obtained with the orthogonal distance regression method, and the red dashed lines 
mark the $\pm$3 $\sigma$ vertical scatter around the best-fit relation.\
{\it Panel c}: age-mass diagram of star cluster candidates in VCC848.\ The masses are stellar masses at birth.\
The results from stellar population fitting with a Gaussian prior constraint on metallicities are shown as filled black circles, while the results  
from fitting without prior constraint on metallicities are shown as red open circles.\ The 90\% completeness limit of stellar mass is 
plotted as blue solid curve.\ The linear relation between maximum cluster mass and age expected from the size-of-sample effect, together 
with a constant cluster formation rate and a canonical cluster mass function of $dN/dM$ $\propto$ $M^{-2}$ is plotted as a blue dashed line, 
which is arbitrarily shifted to match the overall upper mass envelop at intermediate ages.\ The black tilted dotted line represents the cluster 
dissolution limit given by \cite{baumgardt03} for a fiducial galactocentric radius of 0.5 kpc and galaxy rotation velocity of 40 km/s.\ 
{\it Panel d}: average formation rate (Gyr$^{-1}$) of star clusters with birth mass M$_{\star}$ $>$ 10$^{5.25}$ $M_{\odot}$ over the age intervals 
of $0-1$ Gyr and $1-13$ Gyr.
\label{fig_agemass}}
\end{figure*}

The two most massive young star clusters ($<$ 1 Gyr), which have ages $\gtrsim$ $0.3$ Gyr and M$_{\star}$ $\simeq$ 10$^{5.4}$ $M_{\odot}$, are   
near the galactic center, while the youngest star clusters ($\lesssim$ 30 Myr), which have M$_{\star}$ $<$ 10$^{5.0}$ $M_{\odot}$, are at large distances 
($\geq$ 10\arcsec) from the center (Figure \ref{fig_agemass}(a)).\ This suggests that the most intense merging-induced star formation happened 
near the galactic center a few hundred Myr ago, after which active star-forming activities have shifted to larger radii.\ Star clusters with ages $>$ 1 Gyr 
appear to avoid the central disk region and might be regarded as (globular) clusters formed in the progenitor galaxies.\ 

\section{Summary and discussion}\label{sec:sum}
We present the first study of a gas-dominated BCD galaxy (VCC848) which hosts extended stellar shells and thus is confirmed to be a dwarf-dwarf merger.\ 
The stellar light distribution implies a nearly radial encounter between the progenitor galaxies.\ Our isophotal analysis implies a primary-to-secondary stellar 
mass ratio $\lesssim$ 5 and a corresponding dark matter halo mass ratio $\lesssim$ 2 for this advanced major merger.\ We use the age-mass distribution 
of photometrically-selected star cluster candidates to probe the CFR, and find that the average CFR in the past $\sim$ 1 Gyr, which is comparable to the timescales 
expected for dwarf merging events \citep[e.g.][]{bekki08}, was enhanced by a factor of $\sim$ $7-10$.\

The period of most intense merging-induced starburst, which was concentrated near the galactic center a few hundred Myr ago, is expected to be over by now, 
because the secondary progenitor has been almost entirely disrupted.\ The current star formation activities traced by the youngest star clusters mainly occur at large 
galactocentric distances ($\gtrsim$ 1 kpc).\ The overall trend of the temporal and spatial variation of star formation activities in VCC848 is in general agreement with 
the \cite{bekki08} simulations.\ In a companion paper, we will present interferometric observations of the \HI~gas and numerical simulations in order to gain further 
insight into this unique merging system.\

\begin{acknowledgements}
We thank the anonymous referee for very helpful suggestions that improved the manuscript.\
HXZ acknowledges support from the National Key R\&D Program of China (2017YFA0402702), the NSFC grant (Nos.\ 11421303 and 11973039), and the 
CAS Pioneer Hundred Talents Program.\ S.P. acknowledges support from the New Researcher Program (Shinjin grant No.\ 2019R1C1C1009600) through 
the National Research Foundation of Korea.

\end{acknowledgements}



\end{document}